\documentclass[showpacs,amsmath,amssymb,10pt]{article}
\usepackage{graphicx,color}
\usepackage{amsmath}
\usepackage{amssymb}
\usepackage{amsmath,amsfonts,latexsym,graphicx,amssymb}
\usepackage{amsfonts}
\usepackage{amssymb}
\usepackage{mathrsfs}
\usepackage{amsthm}
\usepackage[utf8]{inputenc}
\usepackage{graphicx}
\usepackage{amsmath}
\newcommand{\ot}{{\,\otimes\,}}
\newcommand{{\Cd}}{{\mathbb{C}^d}}
\newcommand{{\Rn}}{{\mathbb{R}^n}}
\newcommand{{\CM}}{{\mathbb{C}^M}}
\newcommand{{\CN}}{{\mathbb{C}^N}}

\def\oper{{\mathchoice{\rm 1\mskip-4mu l}{\rm 1\mskip-4mu l}%
{\rm 1\mskip-4.5mu l}{\rm 1\mskip-5mu l}}}
\def\<{\langle}
\def\>{\rangle}
\newtheorem{thm}{Theorem}

\newtheorem{definition}{Definition}

\newtheorem{remark}{Remark}
\date{}

\begin{document}
\title{\textbf{On separable decompositions of quantum states with strong positive partial transposes}}
\author{B. Bylicka, D. Chru\'sci\'nski, and J. Jurkowski\\
Institute of Physics, Faculty of Physics, Astronomy and Informatics \\
 Nicolaus Copernicus University \\
Grudziadzka 5, 87--100 Toru\'n, Poland}

\maketitle

\begin{abstract}
We analyze a class of positive partial transpose states (PPT) such that the positivity of its partial transposition
is recognized with respect to canonical factorization of the original density operator (Cholesky block decomposition). We call such PPT states  strong PPT states (SPPT). This property, contrary to PPT,  is basis dependent. It is shown that there exists a proper subset of SPPT states which are separable and provide a separable decomposition for any of these states.
\end{abstract}




\section{Introduction}

Quantum entanglement is one of the most remarkable features of
quantum mechanics and it leads to powerful applications like
quantum cryptography, dense coding and quantum computing
\cite{QIT,HHHH}. One of the central problems in the theory of quantum entanglement
is to discriminate between separable and entangled states of composite quantum systems.  Let us
recall that a state represented by a density operator $\rho$
living in the Hilbert space $\mathcal{H}_A \ot \mathcal{H}_B$ is
separable if and only if $\rho$ is a convex combination of product states,
that is,
\begin{equation}\label{SEP}
\rho = \sum_k p_k \, \rho^{(A)}_k \ot \rho^{(B)}_k\ ,
\end{equation}
with $\{p_k\}$ being a probability distribution, and $\rho^{(A)}_k, \
\rho^{(B)}_k$ are density operators  of subsystems $A$ and $B$, respectively \cite{Werner}.
Despite its simplicity the above definition does not tell us when such separable decomposition does exist. It should be stressed that separable decomposition, if it exists, is highly not unique. Moreover, knowing that $\rho$ is separable it is in general very hard to find even one separable decomposition (\ref{SEP}). Therefore, the discrimination between separable and entangled (not separable) states  is usually performed without looking for any specific decomposition. Instead, there are several operational criteria which enable
one to detect quantum entanglement (see e.g.
\cite{HHHH} for the recent review). The most famous
Peres-Horodecki criterion \cite{Peres,Pawel} is based on the partial
transposition: if a state $\rho$ is separable then its partial
transposition $\rho^\Gamma = ({\rm T} \ot \oper)\rho$ is
positive (such states are called PPT state). The structure of this
set is of primary importance in quantum information theory. In
particular it is crucial to understand which PPT states are
separable and which are entangled. It would be very interesting to
find a construction of PPT states which does guarantee
separability. Such construction would shed new light on the basic
structure of PPT states. This is a basic motivation of our paper.

In what follows we use the following convention: let ${\rm dim}\mathcal{H}_A = M$ and ${\rm dim}\mathcal{H}_B = N$. We call a composed system living in $\mathcal{H}_A \ot \mathcal{H}_B$ --- $M \ot N$ system. Let $e_1=|1\>,\ldots,e_M=|M\>$ be an arbitrary orthonormal basis in $\mathcal{H}_A$. Any $M \ot N$ density operator $\rho$ may be represented as follows
\begin{equation}\label{B}
    \rho = \sum_{i,j=1}^M |i\>\<j| \ot \rho_{ij} \ ,
\end{equation}
where $\rho_{ij}$ are operators in $\mathcal{H}_B$, that is, $\rho$ is represented as block $M \times M$ matrix with $N \times N$ blocks $\rho_{ij}$. Performing partial transposition with respect to the basis $|k\>$ in $\mathcal{H}_A$ one obtains
\begin{equation}\label{TB}
    \rho^\Gamma = \sum_{i,j=1}^M |i\>\<j|^{\rm T} \ot \rho_{ij} = \sum_{i,j=1}^M |j\>\<i| \ot \rho_{ij}\ .
\end{equation}
We stress that $\rho^\Gamma$ is basis dependent, however, if $\rho^\Gamma \geq 0$, then this property does not depend on a basis we use for block decompositions (\ref{B}) and  (\ref{TB}). Hence, the notion of a PPT state is basis independent. In this paper we analyze a subclass of PPT states -- Strong Positive Partially Transposed (SPPT) states -- which, contrary to PPT states, are basis dependent. Usually, in physics we prefer notions which are basis or observer independent. We stress that in the case of SPPT states (or rather SPPT block matrices) it is not a drawback but the very essence of the construction. Note, that in practice we usually analyze not an abstract basis independent operator but a basis dependent density matrix.

\section{$2 \ot N$ SPPT states}


To illustrate our construction let us start with $2\ot N$ system (such systems were carefully investigated in \cite{N=2}). Let us fix an orthonormal basis $e_1=|1\>,e_2=|2\>$ in $\mathcal{H}_A$ and introduce the following  upper triangular block matrices $\bf X$ and $\mathbf{Y}$:
\begin{equation}\label{X}
\mathbf{X} = \left( \begin{array}{c|c} X_1 & SX_1  \\ \hline
  \mathbb{O} & X_2  \end{array} \right)\ , \ \ \ \
  \mathbf{Y} = \left( \begin{array}{c|c} X_1 & S^\dagger X_1  \\ \hline
  \mathbb{O} & X_2  \end{array} \right)\ ,
\end{equation}
with arbitrary $N \times N$ matrices $X_1,X_2$ and $S$, and $\mathbb{O}$ denotes an operator in $\mathcal{H}_B$ with vanishing matrix elements, i.e. $\mathbb{O}_{mn}=0$.  Define (unnormalized) density matrix $\rho =  \mathbf{X}^\dagger \mathbf{X}$. One finds \begin{equation}\label{}
    \rho =  \left( \begin{array}{c|c} X_1^\dagger  X_1 & X_1^\dagger  S X_1  \\
\hline  X_1^\dagger  S^\dagger  X_1 & X_1^\dagger  S^\dagger  S
X_1 + X_2^\dagger  X_2
\end{array} \right) \ , \ \ \ \
\rho^\Gamma =
\left( \begin{array}{c|c} X_1^\dagger  X_1 & X_1^\dagger  S^\dagger  X_1  \\
\hline  X_1^\dagger S X_1 & X_1^\dagger  S^\dagger  S X_1 +
X_2^\dagger X_2
\end{array} \right) \ .
\end{equation}
\begin{definition}[\cite{SPPT}]
One says that $\rho$ is SPPT (with respect to $\{e_1,e_2\}$) iff
\begin{equation}\label{}
    \rho^\Gamma = \mathbf{Y}^\dagger \mathbf{Y}\ ,
\end{equation}
that is, iff the following condition is satisfied
\begin{equation}\label{XSSX}
    X_1^\dagger  S^\dagger  S X_1 = X_1^\dagger  S S^\dagger X_1\ .
\end{equation}
\end{definition}

\begin{remark}
A sufficient condition for $\rho$ to be SPPT is that $S$ is normal, i.e. $S^\dagger S = S S^\dagger$. In a recent paper \cite{Chiny} such state was called super SPPT (SSPPT). One has a chain of obvious implications: $\ SSPPT \Rightarrow SPPT \Rightarrow PPT$.
\end{remark}

\begin{remark}
If $X_1$ has a full rank (${\rm rank}\, X_1 = N$), then formula (\ref{XSSX}) implies that $S$ is normal.
\end{remark}

\begin{remark}
If $\rho$ is SPPT with respect to $\{e_1,e_2\}$ it needs not be SPPT with respect to another basis $\{f_1,f_2\}$. Consider for example
\begin{equation}\label{}
    X_1 = \left(\begin{array}{cc}
    2 & 1 \\
    1 & -1 \end{array}\right),\qquad S=\left(\begin{array}{cc}
    0 & 1 \\
    -1 & 0 \end{array}\right),\qquad X_2=\left(\begin{array}{cc}
    1 & 0 \\
    0 & 1 \end{array}\right).
\end{equation}
It is SPPT since $S$ is normal. However, if we perform a local unitary transformation
\begin{equation}\label{}
\rho \ \rightarrow \    (U \ot \mathbb{I}) \rho (U^\dagger \ot \mathbb{I})\ ,
\end{equation}
with $U$ being a Hadamard gate
$$   U = \frac{1}{\sqrt{2}}\, \left( \begin{array}{cc} 1 & 1  \\ 1 & -1  \end{array} \right) \ , $$
the transformed $\rho$ is no longer SPPT. In a recent paper \cite{Chiny} authors declare that SPPT are basis independent. However, the proof in \cite{Chiny} is not correct.
\end{remark}

In a recent paper \cite{Ha2} Ha shows that if $N\leq 4$ any SPPT state is separable. However, for $N\geq 5$ there are SPPT entangled states.
Interestingly, for arbitrary $N$ one has the following
\begin{thm}
If $\rho$ is super SPPT, then it is separable.
\end{thm}

Proof: we follow the same idea as in \cite{Ha2}. One has a natural decomposition
\begin{equation}\label{12}
    \rho = \left( \begin{array}{c|c} X_1^\dagger  X_1 & X_1^\dagger  S X_1  \\
\hline  X_1^\dagger  S^\dagger  X_1 & X_1^\dagger  S^\dagger  S
X_1  \end{array} \right)  +
\left( \begin{array}{c|c} \mathbb{O}  & \mathbb{O} \\
\hline  \mathbb{O} &  X_2^\dagger  X_2
\end{array} \right) \ .
\end{equation}
The second term is obviously separable being a product $|2\>\<2| \ot X_2^\dagger  X_2$. Concerning the first term observe that since $S$ is normal it provides a spectral decomposition $S = \sum_{k=1}^2 \lambda_k P_k$ with complex $\lambda_k$ and rank-1 projectors $P_k$.  Hence, it may be rewritten as follows
\begin{equation}\label{1}
\left( \begin{array}{c|c} X_1^\dagger & \mathbb{O}  \\
\hline  \mathbb{O} & X_1^\dagger \end{array} \right)  \left( \begin{array}{c|c} \mathbb{I} & S   \\
\hline S^\dagger  & S^\dagger  S   \end{array} \right) \left( \begin{array}{c|c} X_1 & \mathbb{O}  \\
\hline  \mathbb{O} & X_1 \end{array} \right) = \sum_{k=1}^2 \sigma_k \ot X_1^\dagger P_k X_1 \ ,
\end{equation}
where
\begin{equation}\label{}
    \sigma_k = \left( \begin{array}{cc} 1 & \lambda_k \\ \overline{\lambda}_k & |\lambda_k|^2 \end{array} \right) \ .
\end{equation}
Note that $\sigma_k = |\psi_k\>\<\psi_k|$ with $|\psi_k\> = |1\> + \overline{\lambda}_k |2\>$ which proves that $\sigma_k \geq 0$ and hence $\rho$ is separable. \hfill $\Box$

\section{$M \ot N$ SPPT states}

In the general case a state $\rho$  may be considered as an $M \times M$ matrix with entries being operators from $\mathfrak{B}(\mathcal{H}_B)$. Positivity of $\rho$ implies that $\rho = {\bf X}^\dagger {\bf X}$.  Let us consider the following class of upper triangular
block matrices $\bf X$
\begin{equation}\label{XN}
{\bf X} = \left( \begin{array}{c|c|c|c|c} X_1 & S_{12}X_1 & S_{13}X_1 & \ldots & S_{1M}X_1 \\
\hline  \mathbb{O} & X_2 & S_{23}X_2 &\ldots & S_{2M}X_2 \\ \hline \vdots &
\vdots & \ddots & \vdots & \vdots \\ \hline \mathbb{O} & \mathbb{O} & \mathbb{O} & X_{M-1} &
S_{M-1,M}X_{M-1} \\ \hline \mathbb{O} & \mathbb{O} & \mathbb{O} & \mathbb{O} & X_M
\end{array} \right) \ ,
\end{equation}
where $X_k$ and $S_{ij}\ (i<j)$ belong to $\mathfrak{B}(\mathcal{H}_B)$. This block matrix may be written in a compact form
\begin{equation}\label{}
    \mathbf{X} = \sum_{i,j=1}^M |i\>\<j| \ot \mathbf{X}_{ij}\ ,
\end{equation}
with $\mathbf{X}_{ij} = S_{ij} X_{i}$, where we assume that $S_{ii} = \mathbb{I}$ and $S_{ij} = \mathbb{O}$ for $i>j$.
One has
\begin{equation}\label{}
    \rho = \mathbf{X}^\dagger \mathbf{X} = \sum_{i,j=1}^M |i\>\<j| \ot \rho_{ij}\ ,
\end{equation}
where the blocks are defined by
\begin{equation}\label{A}
    \rho_{ij} = \sum_{k=1}^{i} X_k^\dagger  S^\dagger_{ki} S_{kj} X_k    \ ,
\end{equation}
for $i \leq j$, and for partially transposed block matrix
\begin{equation}\label{A1}
    \rho^\Gamma =  \sum_{i,j=1}^M |j\>\<i| \ot \rho_{ij} = \sum_{i,j=1}^M |i\>\<j| \ot \widetilde{\rho}_{ij}\ ,
\end{equation}
with
\begin{equation}\label{}
    \widetilde{\rho}_{ij} = \rho_{ji} = \sum_{k=1}^{i} X_k^\dagger  S^\dagger_{kj} S_{ki} X_k   \ ,
\end{equation}
for $i \leq j$ (one obviously has $\rho_{ij} = \rho_{ji}^\dagger$). Now, in analogy to $2 \ot N$ case we have the following

\begin{definition}
$\rho$ is SPPT (with respect to $\{e_1,\ldots,e_M\}$) if
 $\rho^{\Gamma} = {\bf Y}^\dagger {\bf Y}$ where ${\bf Y}$ is given by (\ref{XN}) with $S_{ij}$ replaced by $S_{ij}^\dagger$.
\end{definition}

One easily finds that $\rho$ is SPPT if
\begin{equation}\label{}
\sum_{k=1}^{i} X_k^\dagger S_{kj}^\dagger S_{ki} X_k =
\sum_{k=1}^{i} X_k^\dagger S_{ki} S^\dagger_{kj} X_k\ , \ \ \ \  i \leq j\ .
\end{equation}
In particular the above conditions are satisfied if
\begin{equation}\label{SS-SS}
S_{ki} S^\dagger _{kj} = S^\dagger _{kj} S_{ki}\ ,
\end{equation}
for $k < i \leq j $. Following \cite{Chiny} we call SPPT states satisfying (\ref{SS-SS}) super SPPT.

\begin{thm}
A super SPPT state  is separable.
\end{thm}

Proof: it is clear from (\ref{A}) and (\ref{A1}) that
\begin{equation}\label{+++}
    \rho = \rho_1 + \rho_2 + \ldots + \rho_M\ ,
\end{equation}
where
\begin{equation}
\rho_k   = \sum_{i,j=k}^M  |i\> \<j| \ot X_k^{\dagger} S_{ki}^{\dagger} S_{kj} X_k\ .
\end{equation}
Note that the sum starts with $i,j=k$  due to the fact that $S_{ki} = \mathbb{O}$ for $i<k$.
We show that all $\rho_k$ are separable and hence (\ref{+++}) provides separable decomposition of $\rho$. Condition (\ref{SS-SS}) implies that  $S_{ki}$ for $k<i$ defines a family of normal and mutually commuting operators. Hence
\begin{equation}
S_{ki}=\sum_{l=1}^N \lambda_l^{(ki)} P^{(k)}_l \ ,
\end{equation}
where $\lambda_l^{(ki)}$ are complex and  $P^{(k)}_l$ are rank-1 projectors. It gives therefore
\begin{equation}
\rho_k=\sum_{l=k}^M \sigma^{(k)}_l \ot X_k^{\dagger} P^k_l X_k,
\end{equation}
where $\sigma^{(k)}_l=\sum_{i,j=k}^N \lambda^{(ki)}_l \overline{\lambda^{(kj)}_l} |i\>\< j|$. Note, that $\sigma^{(k)}_l = |\psi^{(k)}_l\>\<\psi^{(k)}_l|$ with $|\psi^{(k)}_l\> = \sum_{i=k}^N \overline{\lambda^{(ki)}}_l |i\>$ which proves that $\sigma^{(k)}_l$ are positive operators and hence $\rho_k$ is separable.  \hfill $\Box$

\begin{remark} In \cite{SPPT} it was conjectured that all SPPT states are separable. This statement is not true as was observed by Ha \cite{Ha1} who provided an example of $3 \ot 3$ entangled SPPT state.
\end{remark}

An interesting class of super SPPT states is provided by the so called {\em classical-quantum states} (CQ)
\begin{equation}\label{CQ}
    \rho = \sum_n p_n |e_n\>\<e_n| \ot \sigma_n\ ,
\end{equation}
where $|e_n\>$ defines an orthonormal basis in $\mathcal{H}_A$, $\sigma_n$ are density operators in $\mathcal{H}_B$, and $p_n$ is a probability distribution. Such states have vanishing quantum discord and were recently intensively investigated (see recent review \cite{Discord}).
It was shown in Ref. \cite{Bogna} that $2 \ot N$ states with vanishing discord are super SPPT with respect to an arbitrary basis in the qubit Hilbert space. However, it is lo longer true for $M \ot N$ states with $M >2$ (actually, authors of Ref. \cite{Chiny} provided a proof that this statement is true but  their proof is not correct). A subclass of CQ defines so called {\em classical-classical states } (CC), i.e. states for which $\sigma_n$ are mutually commuting. It implies that there exists an orthonormal basis $|f_m\>$ in $\mathcal{H}_B$ such that
\begin{equation}\label{CC}
    \rho = \sum_n p_{nm} |e_n\>\<e_n| \ot |f_m\>\<f_m|\ .
\end{equation}
CC states are, therefore, fully characterized by the classical joint probability distribution $p_{nm}$. Let us observe that a CC state rewritten in another basis $|\widetilde{e}_n\>$ in $\mathcal{H}_A$ has the following form $\rho = \sum_{n,m} |\widetilde{e}_n\>\<\widetilde{e}_m| \ot \rho_{nm}$ and the blocks $\rho_{nm}$ are diagonal in the basis   $|f_m\>$. It is therefore clear that a CC state is super SPPT with respect to an arbitrary basis in $\mathcal{H}_A$.  For a recent discussion on quantifying classical and quantum correlations see
the series of papers in \cite{IJMPB}. CQ and CC states and the corresponding CQ and CC quantum channels have been  recently analyzed in \cite{CQ1,CQ2}.

\section{Conclusions}

We analyzed a class of SPPT states in $\mathcal{H}_A \ot \mathcal{H}_B$ with respect to a fixed orthonormal basis $\{e_1,\ldots,e_M\}$ in $\mathcal{H}_A$ with $M = {\rm dim}\, \mathcal{H}_A$. We stress that a property to be SPPT state is always defined with respect to a fixed basis and hence it is basis dependent.  It is not a drawback but the very essence of the construction. In particular we showed that a state which is super SPPT is separable. Moreover, we provided separable decomposition for any super SPPT state.

Now, any SPPT state is PPT and any super SPPT is separable. One may ask whether the converse statements are also true. Interestingly, there are PPT states which are not SPPT with respect to any basis. Consider for example PPT Werner states for two qubits represented in the computational basis as
\begin{equation}\label{}
    W_p = \frac{1}{6}\left(\begin{array}{cc|cc}
    2p & 0 & 0 & 0 \\
    0 & 3-2p & 4p-3 & 0 \\ \hline
    0 & 4p-3 & 3-2p & 0 \\
    0 & 0 & 0 & 2p \end{array}\right),\qquad  0 \leq p\leq 1\,.
\end{equation}
It is PPT (and hence separable) for $p \geq \frac 12 $. One easily finds
\begin{equation}\label{}
    X_1 = \left(\begin{array}{cc}
    \sqrt{\frac{p}{3}} & 0 \\
    0 & \sqrt{\frac{3-2p}{6}} \end{array}\right) , \quad
    S= \left(\begin{array}{cc}
    0 & 0 \\
    \frac{4p-3}{\sqrt{2p(3-2p)}} & 0 \end{array}\right)\ , \quad
    X_2 = \left(\begin{array}{cc}
    \sqrt{\frac{2p(p-1)}{2p-3}} & 0 \\
    0 & \sqrt{\frac{p}{3}} \end{array}\right)
\end{equation}
which shows that $W_p$ is not SPPT for $p\neq \frac{3}{4}$ ($W_{\frac 34} = \frac 14 \mathbb{I}\ot \mathbb{I}$ and hence it is SPPT in all basis). Now, according to the defining property of the Werner state one has
\begin{equation}\label{}
    U \ot U W_p U \ot U = W_p\ ,
\end{equation}
for each unitary operator in $\mathbb{C}^2$. This invariance shows that $W_p$ is not SPPT in the transformed basis $\{U|1\>,U|2\>\}$.
It would be interesting to find a class of states with the following property:   if $\rho$ is PPT (separable), then there exists an orthonormal basis $\{e^*_1,\ldots,e^*_M\}$ in $\mathcal{H}_A$ such that $\rho$ is SPPT (super SPPT) with respect to this basis. Note that this class of separable states provides natural separable decomposition. Finally, it would be interesting to generalize SPPT states for multipartite setting.

\section*{Acknowledgements}

BB was supported by the grant  2011/01/N/ST2/00393. DC and JJ were supported by the National Science Center project  
DEC-2011/03/B/ST2/00136.

\end{document}